\documentstyle[sprocl]{article}

\input{psfig}
\def\etal{\it et al.\rm}
\def\xo{$X_0$\ }
\def\be{\begin{equation}}
\def\ee{\end{equation}}
\def\bea{\begin{eqnarray}}
\def\eea{\end{eqnarray}}

\begin{document}

{\hfill LBNL - 42069}
\vskip .01 in
{\hfill August, 1998}
\vskip .2 in

\title{The LPM Effect: Comparing SLAC E-146 Data with Experiment}

\author{Spencer R. Klein}

\address{Lawrence Berkeley National Laboratory \\
Berkeley, CA, 94720, USA \\E-mail: srklein@lbl.gov}


\maketitle

\abstracts{ The suppression of photon bremsstrahlung due to a variety
of in-medium effects is discussed.  Different electrodynamic
suppression effects are discussed, and compared with the related color
analogs.  Higher order effects are considered, and found to be
important.  Data from SLAC E-146 is discussed, and compared with
theory.  The effect of finite thickness targets is emphasized, since
nuclear size is such an important limiting factor 
for the chromodynamics effects.}

\centerline{Presented at the \it Fourth Workshop on Quantum
Chromodynamics \rm}
\centerline{June 1-6, 1998, Paris, France.}

\section{Introduction}

Originally, the LPM effect referred to the suppression of electron
brems\-strah\-lung or pair production due to multiple scattering.  More
recently, it has been applied to the effects of the nuclear medium on
quark and gluon interactions.  This subject is of interest because
calculations predict that quark or gluons should radiate increased
energy in traversing a quark gluon plasma, compared to normal nuclear
matter.

This talk will consider the electrodynamic version of the LPM effect.
I will discuss recent experimental results from SLAC E-146, and
compare these results with theory.  This might seem like an odd choice
for a conference devoted to quantum chromodynamics.  However,
electrodynamics can be an important point of comparison for the
chromodynamics calculations.  For electrodynamics, it is possible to
identify a variety of different kinematic regimes, with significantly
different photon spectral indices.  And, the electrodynamics
calculations can be tested experimentally, over a broad range of
target thicknesses.

\section{Suppression Mechanisms}

The LPM effect stems from the formation length, the
distance over which an interaction such as pair production or
bremsstrahlung occurs.  For bremsstrahlung from an isolated atom, this
distance is
\begin{equation}
l_f = {2\hbar E(E-k) \over m^2 k}
\end{equation}
where $E$ is the incoming electron energy, $k$ is the photon energy
and $m$ the electron mass.  This distance can be very long; for a 25
GeV electron emitting a 1 MeV photon, $l_f= 1$ mm!  Classically, if
the electron is disturbed while traversing this distance, then the
emission can be disturbed.  In field theoretical language, other
interactions within $l_f$ can interfere with the bremsstrahlung,
reducing its amplitude.  
In dense media, the Bethe-Heitler
$1/k$ bremsstrahlung photon spectrum is suppressed by a
factor $S$\ \cite{LP}:
\begin{equation}
S= {\sigma\over\sigma_{BH}} = \sqrt{kE_{LPM} \over E(E-k)}
\end{equation}
where $E_{LPM}= m^4 X_0/E_s^2$ where $X_0$ is the radiation length
of the material, and $E_s=m\sqrt{4\pi/\alpha}$.  

In 1956, Migdal used the Boltzman transport equation to model multiple
scattering, calculating the emission for each path\ \cite{Migdal}.  He
used a simple model for the potential, and found results good to
logarithmic accuracy.

One limitation of Migdals result is that neglected surface effects,
which are important for targets of finite thickness.  This is
important for QCD, where the target size is limited to a nuclear
diameter.  Gol'dman\ \cite{goldman} and Ternovskii\ \cite{ternovskii}
extended Migdals calculation to include finite thickness targets.  In
the limit $T\ll l_f$, the Bethe-Heitler $1/k$ spectrum is recovered,
albeit at a reduced intensity, proportional to $\ln{T}$. Since E-146,
there have been several new calculations\ \cite{newwork}; since
several of the authors are speaking here, I will not further discuss
them further.

One interesting aspect of the QED is that it allows for a variety of
different suppression mechanisms.  In dielectric suppression\
\cite{dprl}, the produced photon interacts with the electrons in the
medium.  This bulk interaction, mediated by forward Compton
scattering, is best expressed in terms of the dielectric constant of
the medium, $\epsilon(k) = 1 - (\hbar\omega_p)^2/k^2$, where
$\omega_p$ is the plasma frequency.  This interaction gives the photon
an effective mass $\omega_pc^2$, which shortens the formation length,
and produces a suppression that scales as $k^2$:
\begin{equation}
S =  { k^2 \over k^2 + (\gamma\hbar\omega_p)^2 }.
\label{sdiel}
\end{equation}
where $\gamma=E/m$. A similar effect can occur when a radiated gluon
undergoes further interactions.  Although these interactions are
included in current calculations, the specific effects of these
diagrams have not been considered separately.

Another mechanism occurs when the $l_f$ is longer than the radiation
length.  Then, the nascent photon can interact before it is fully
created.  This limits $l_f$ to $X_0$, suppressing photon emission\
\cite{myreview}.  Similarly, bremsstrahlung can also suppress pair
production when a produced leptons radiates a photon in the formation
zone.  For electrodynamics, this effect only occurs at extremely high
energies.  For QCD however, the interaction length inside a nucleus
can be smaller than $l_f$, so multiple interactions in a single
formation length are likely.  This higher order effect is not
considered in current calculations. Unfortunately, this 'correction'
is likely to be very large, and numerical predictions of quark $dE/dx$
in nuclear media should be used with great caution.  This problem will
greatly complicate the interpretation of energy loss measurements
planned for RHIC.

Suppression can also occur when bremsstrahlung or pair production
occurs in an external magnetic field\ \cite{myreview}.  In the absence
of a bulk color magnetic field, this effect is unlikely to be
important in QCD. These different suppression mechanisms are
summarized in Table 1.

\begin{table}[t]
\caption{Bremsstrahlung photon spectral indices.}
\vspace{0.2cm}
\begin{center}
\footnotesize
\begin{tabular}{|c|c|c|l|}
\hline
Region & Dominant Mechanism & Photon Spectrum & Importance in QCD \\
\hline
       &       &           &      \\
none   &   -   &  $k^{-1}$ &   ?  \\
LPM    &  Multiple Scattering & $k^{-1/2}$ &  yes \\
Pair Production & Pair Production & $k^0$  & very \\
Dielectric  &  Compton Scattering & $k$  & ? \\
Magnetic        & Magnetic Field  &  $k^{-1/3}$ & no \\
\hline
\end{tabular}
\end{center}
\end{table}

\section{E-146 Data and Analysis}

\begin{figure}
\centerline{\psfig{file=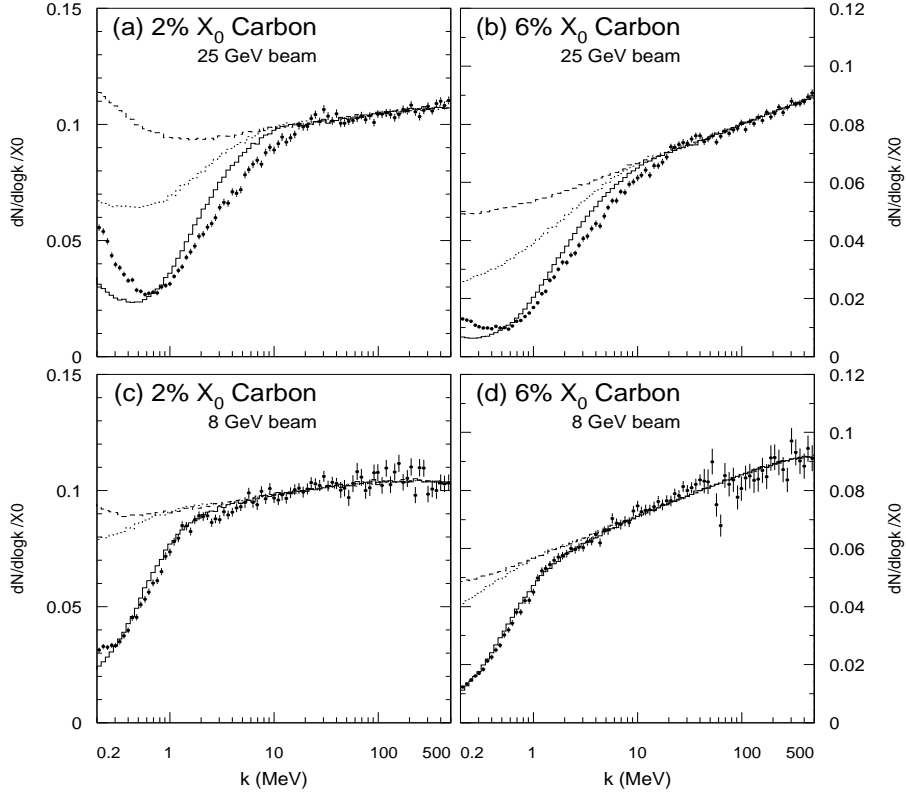,height=4.15in,width=4.75in,%
bbllx=30pt,bblly=165pt,bburx=570pt,bbury=670pt}}
\caption{Data from SLAC-E-146 compared with MC predictions for
200 keV to 500 MeV photons from 8 and 25 GeV electrons passing through
2\% and 6\% $X_0$ carbon targets. The cross sections are given as
$dN/d(\log k)/X_0$ where $N$ is the number of events per photon energy
bin per incident electron, for (a) 2\% \xo\ carbon and (b) 6\% \xo\
carbon targets in 25 GeV electron beams, while (c) shows the 2\% \xo\
carbon and (d) the 6\% \xo\ carbon target in an 8 GeV beam.  Three
Monte Carlo curves are shown.  The solid line includes LPM and
dielectric suppression of bremsstrahlung, plus conventional transition
radiation.  Also shown are the Bethe-Heitler plus transition radiation
MC (dashed line) and LPM suppression only plus transition radiation
(dotted line).  Adapted from Anthony \etal\ (1997).}
\label{carbon}
\end{figure}

The SLAC E-146 collaboration has studied LPM\ \cite{prl}\ \cite{prd} and
dielectric suppression\ \cite{dprl} by observing 200 keV to
500 MeV photons produced by 8 and 25 GeV electrons passing through a
variety of targets.  For most materials, two different thickness
targets were studied.  Since the experiment is well described
elsewhere, here I will focus on the data and its implications for
theory.  The photon flux is histogrammed in logarithmic bins in $k$,
$1/X_0 dN/d\log{k}$.  This binning allowed the histograms to cover
many orders of magnitude in $k$.  It also flattened out the $1/k$
Bethe-Heitler spectrum.

Figures 1-3 show the E-146 data for carbon, uranium and thin gold
targets.  These targets cover a wide range in density, and also in
$l_f/T$.  Figure 1 compares the carbon ($T\gg l_f$) data with
predictions based on Bethe-Heitler, LPM suppression only, and LPM plus
dielectric suppression; both mechanisms are clearly required to match
the data.  However, the 25 GeV data shows a significant disagreement
in the region $k\approx E^2/E_{LPM}$.  Figure 2 shows uranium data,
compared with curves based on LPM plus dielectric suppression.  The
data and theory agree for $k>10$ MeV, but the data rises above the
theory for $k<10$ MeV.  The difference can be attributed to the small
thickness of the targets.  For these target thicknesses, when $k< 10$
MeV, $T\approx l_f$, so surface effects are likely to be important.
Several early calculations of the surface effects, shown in the
figure, fail to reproduce the data.  Newer calculations do appear to
reproduce the surface terms, but are not easily comparable with data
because they do not localize the emission, and hence cannot be easily
included in a simulation.

\begin{figure}
\centerline{\psfig{file=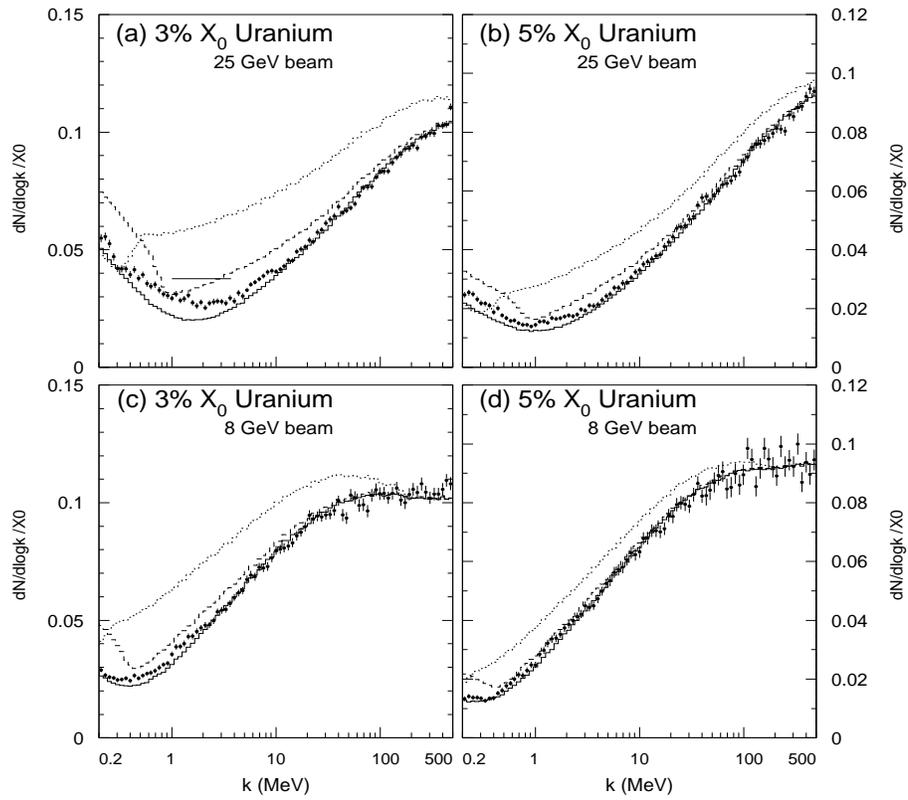,height=4.15in,width=4.75in,%
bbllx=30pt,bblly=165pt,bburx=570pt,bbury=670pt}}
\caption{SLAC E-146 data and Monte Carlo for 3\% \xo\ and 5\% \xo\
uranium targets in 8 and 25 GeV electron beams.  The solid line shows
the LPM and dielectric suppression, conventional transition radiation
Monte Carlo prediction. The other lines include simulations based on
calculations of transition radiation due to Pafomov (dashed line) and
Ternovskii (dotted line).  Adapted from Anthony \etal\ (1997).}
\label{uranium}
\end{figure}

Figure 3 shows the E-146 data for electrons passing through a 0.7 \xo
(23$\mu$m) thick gold target.  Here, $l_f>T$ for $k< 7$ MeV.  In this
regime, the target interacts as a coherent whole, and the
Bethe-Heitler $1/k$ spectrum is recovered, albeit at a reduced
intensity\cite{ternovskii}.  This flattening is also predicted by
newer calculations, such as those by Blankenbecler and Drell.  In this
case, the target is thin enough that multiple interactions by a single
electron are unlikely, and so the calculation can be directly compared
with the data.

\begin{figure}
\centerline{\psfig{file=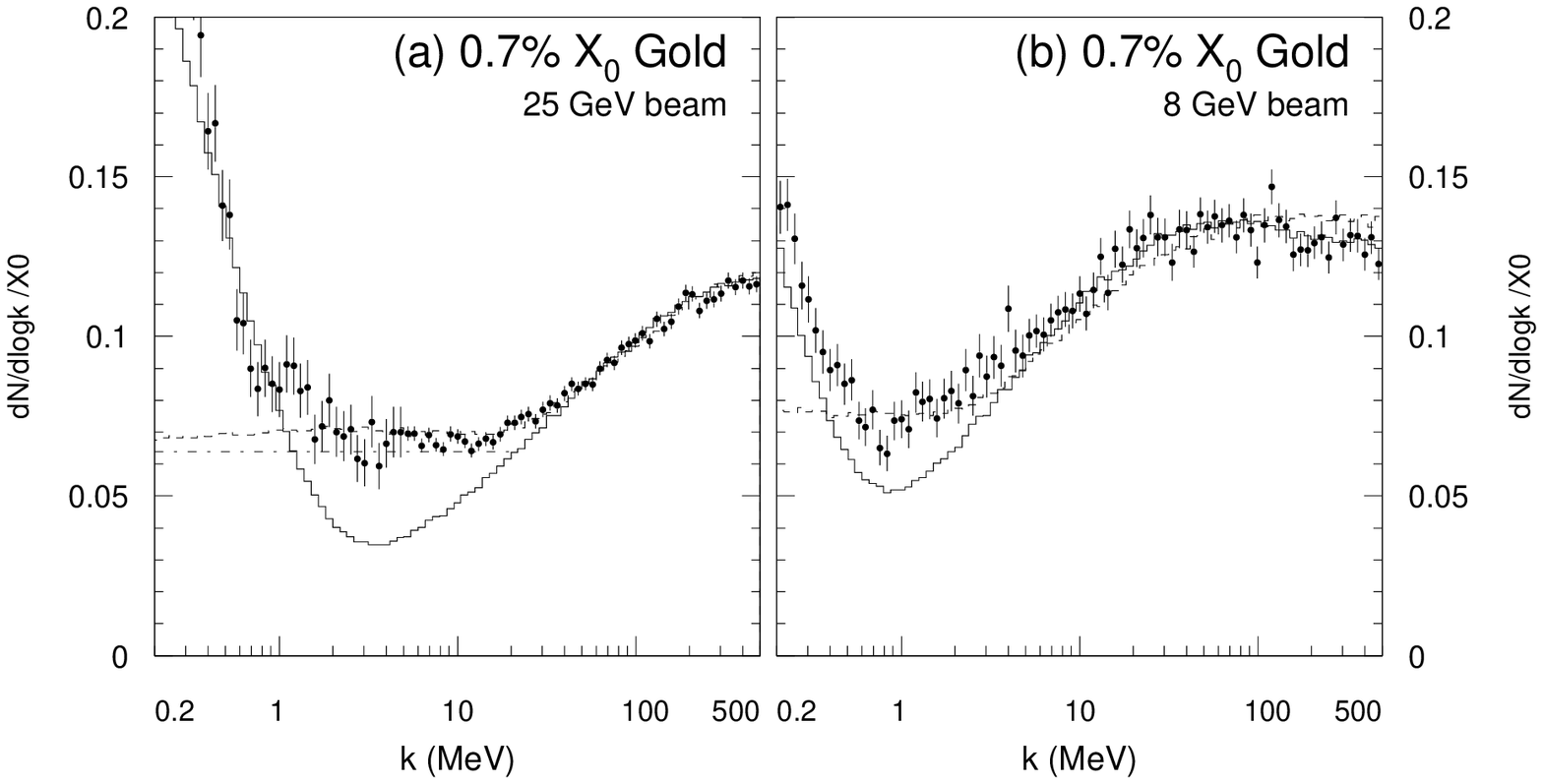,height=2.2in,width=4.75in,%
bbllx=30pt,bblly=160pt,bburx=572pt,bbury=436pt}}
\caption{SLAC E-146 data on 8 and 25 GeV electrons hitting a
0.7\%~$X_0$ gold target.  Shown are calculations by Blankenbecler and
Drell (dashed line), and Shulga and Fomin (dot-dashed line). For
comparison, the Migdal MC is shown as the usual solid line.  Adapted
from Anthony \etal\ (1997).}
\label{gold}
\end{figure}

The systematic error for these measurements are small, ranging from
3.3\% at the higher photon energies, up to 17\% for $k< 5$ MeV in a 25
GeV beam.  The systematic errors are smaller than the discrepancy
between the carbon data and theory.

\section{Conclusions}

I have discussed several mechanisms which can suppress electron
brems\-strah\-lung.  Different mechanisms are important in different
kinematic regions.  At current accelerators, for electrodynamics,
suppression due to multiple scattering is the most important, followed
by dielectric suppression.

At much higher electron energies, pair production can suppress
brems\-strahlung in the regime $l_f > X_0$.  In this regime, in fact,
the concept of individual interactions in an electromagnetic shower
break down, leaving an (so far) unsolvable complex Feynman diagram
containing multiple steps in a shower.  For colored interactions in
dense media, $X_0$ is much smaller, and these higher order diagrams are
likely to be very important, even at near-future colliders.

The accuracy of the electrodynamics calculations is shown by data from
SLAC E-146.  The data generally matches the theory to within 10\%.
The one exception to this is with the light (low $Z$) targets, where
the data is somewhat below the theory around $k=E(E-k)/E_{LPM}$.  The
reason for this is unknown, but may stem from an inadequate treatment
of atomic effects.

\section*{Acknowledgments}

I would like to thank my E-146 colleagues for their support.  This
work was supported by the U.S. D.O.E. under contract
DE-AC03-76SF00098.

\end{document}